\documentclass{edp-conf}
\usepackage{graphicx}
\begin{document}
\def\la{\buildrel<\over\sim}
\def\ga{\buildrel>\over\sim}

\TitreGlobal{SF2A 2005}

\title{PROPERTIES OF REGIONS FORMING THE FeII EMISSION LINES IN Be STARS}

\author{Arias, M.L.}\address{Facultad de Ciencias Astron\'omicas y 
Geof\'\i sicas, Universidad de La Plata, Argentina}
\author{Zorec, J.}\address{Institut d'Astrophysique de Paris, UMR7095 CNRS, 
Universit\'e Pierre \& Marie Curie}
\author{Cidale, L.$^1$}
\author{Ringuelet, A.$^1$}

\runningtitle{FeII emission lines in Be stars}

\index{Arias, J.}
\index{Zorec, J.}
\index{Cidale, L.}
\index{Ringuelet, A.}

\maketitle

\begin{abstract} We study FeII and Balmer hydrogen emission lines observed 
simultaneously of 18 southern Be stars. We use the self-absorption-curve 
method (SAC) to determine the optical depth regime of FeII emission lines and
to derive first insights on the physical properties of their forming regions. 
\end{abstract}

\section{Introduction}

 While emission in the first Balmer terms appears for all B sub-spectral types, 
in Fe II lines is apparent mainly for types earlier than B5 with rare 
exceptions in latter B sub-types. Many authors carried out more or less 
systematic studies on the strongest Fe II lines in Be stars. However, it has
not been yet stated the optical depth regime that these lines obey in the 
circumstellar envelope (CE) nor the actual location of their formation region.
To this purpose, several members of different multiplets of Fe II lines in the 
optical range are to be analyzed with empirical methods that avoid, as much as
possible, model-dependent diagnostics.

\section{Observations and some correlations}

 The observations of 18 stars were obtained at the Complejo Astron\'omico El 
Leoncito (CASLEO), San Juan, Argentina, on March and September 1996 and March 
2002 using the 2.15 m telescope and a REOSC echelle Cassegrain spectrograph 
with mean resolution R = 11500.\par
 An interesting result found is that the equivalent width of the central 
depression of the profiles of H$\beta$, H$\gamma$ and H$\delta$ correlates 
with $V\!\sin i$. This can be consistent with the material producing the
emission and self absorption in these lines being in flattened enough region 
of the extended envelope interior to the region where H$\gamma$ and H$\delta$
are formed.

\section{The SAC curves and Results}

 The SAC method developed by Friedjung \& Muratorio (1987) and Muratorio \&
Friedjung (1988) makes explicit the opacity effect of the emitting layers on 
the emission lines. It then may carry information on the line opacity.
Horizontal displacements of SAC segments of multiplets with a common lower 
level, as well as vertical displacements of segments relative to multiplets 
with common upper levels, can lead to the estimate of relative level 
populations and to derive their excitation temperature. The comparison of the
empirically re-composed SAC with the SAC function $Q=$ $(1-e^{-\tau})/\tau$, 
where $\tau$ is the opacity of a reference multiplet, may then lead to an 
estimate of the optical depth of lines and to the radius of the line emitting
region. The relative displacements of individual SAC segments of multiplets 
with common upper or lower levels can be expressed as: $|\Delta(X,Y)|=$ 
$(\chi_1-\chi_2)(5040/T_{\rm ex})$, where $\chi_1$ and $\chi_2$ are the 
excitation potentials of two given levels. They can be used to obtain the 
excitation temperature $T_{\rm ex}$. However, when $T_{\rm ex}$ is high the 
displacements are too small to be measured, as it is the case for our program
stars. To obtain an estimate of the excitation temperature, we used then an
alternative iteration process.\par
 SAC curves showed that $\partial Q/\partial\tau <0$ which implies that FeII
are optically thick. The excitation temperatures we obtained range from 4300 
to 13300 K. The radii of the FeII emitting regions range from 4.3 to 1.1$R_*$. 
The obtained radii are systematically smaller than those obtained using 
Huang's (1972) expression, $\Delta_p=2V\sin i(R/R_*)^{-j}$, suited for
optically thin lines. However, as shown by our results, Fe~II lines are 
optically thick and the separation of the peaks is further a function of the 
velocity field and the optical thickness (Cidale \& Ringuelet 1989).

\section{Conclusions}

 We conclude that all Fe~II lines observed in the optical range are optically 
thick and are formed very close to the central star at distances that range 
from 1 to 4.25 stellar radii, where the excitation temperature range from 
4300K to 13300K according to the star. In our case, the SAC method could not 
be used in its whole potentiality and thus, it was not possible to derive the
populations of the levels involved in the different transitions. We consider 
that it is necessary to represent in some detail the physical characteristics 
of the Be CE based on the first order properties derived here. In particular,
it is suited to include the source function of the lines in the SAC and to 
study the influence of the optical depth on the shape of the emission line 
profiles.\par

\end{document}